\documentclass[useAMS,usenatbib]{mn2e}

\usepackage{graphicx}
\newcommand{\chandra}{{\it Chandra}}

\newcommand{\xmm}{{\it XMM-Newton}}

\title[Multifrequency nature of the 0.75 mHz feature in the X-ray light curves of the nova V4743~Sgr]{Multifrequency nature of the 0.75 mHz feature in the X-ray light curves of the nova V4743~Sgr}
\author[A. Dobrotka and J.-U. Ness]{A. Dobrotka$^{1}$\thanks{E-mail: andrej.dobrotka@stuba.sk}, J.-U. Ness$^{2}$\thanks{E-mail: juness@sciops.esa.int}\\
$^{1}$Department of Physics, Institute of Materials Science, Faculty of Materials Science and Technology, Slovak University\\ of Technology in Bratislava, J\'ana Bottu 25, 91724 Trnava, The Slovak Republic\\
$^{2}$European Space Astronomy Centre, PO Box 78, 28691 Villanueva de la Ca\~nada, Madrid, Spain: juness@sciops.esa.int}

\begin{document}

\date{Accepted ???. Received ???; in original form \today}

\pagerange{\pageref{firstpage}--\pageref{lastpage}} \pubyear{2009}

\maketitle

\label{firstpage}

\begin{abstract}
We present timing analyses of eight X-ray light curves and one optical/UV
light curve of the nova V4743\,Sgr (2002) taken by \chandra\ and \xmm\ on
days after outburst: 50 (early hard emission phase), 180, 196, 302, 371,
526 (super soft source, SSS, phase), and 742 and 1286 (quiescent emission
phase). We have studied the multifrequency nature and time evolution of the dominant peak at $\sim 0.75$\,mHz using the standard Lomb-Scargle method and a 2-D sine fitting method. We found a double structure of the peak and its overtone for days 180 and 196.
The two frequencies were closer together on day 196, suggesting that the
difference between the two peaks is gradually decreasing. For
the later observations, only a single frequency can be detected, which is
likely due to the exposure times being shorter than the beat period
between the two peaks, especially if they are moving closer together.
The observations on days 742 and 1286 are long enough to detect two
frequencies with the difference found for day 196, but we confidently find only
a single frequency.
We found significant changes in the oscillation frequency and amplitude.
We have derived blackbody temperatures from the SSS spectra,
and the evolution of changes in frequency and blackbody temperature
suggests that the 0.75-mHz peak was modulated by pulsations.
Later, after nuclear burning had ceased, the signal stabilised at a single
frequency, although the X-ray frequency differs from the optical/UV frequency
obtained consistently from the OM onboard \xmm\ and from ground-based
observations. We believe that the late frequency is the white dwarf rotation
and that the ratio of spin/orbit period strongly supports that the system is
an intermediate polar.
\end{abstract}


\section{Introduction}
\label{intro}

Cataclysmic variables are interacting binary systems, consisting of a white
dwarf primary and a late main sequence star, where accretion takes place
from the cool companion to the white dwarf. If the accretion rate is in the
range between
$(1-4)\times 10^{-7}$\,M$_\odot$\,yr$^{-1}$, steady nuclear burning can
establish, thus the hydrogen content of the accreted material is fused
to helium at the accretion rate upon arrival on the white dwarf surface 
\cite{vandenheuvel1992}. If the accretion rate is
$<10^{-7}$\,M$_\odot$\,yr$^{-1}$ then the hydrogen-rich material settles
on the white dwarf surface and ignites explosively in a thermonuclear
runaway, after a critical amount of mass of
$10^{-6}-10^{-4}$\,M$_\odot$ (depending on white dwarf mass) has been
reached. Such outbursts are commonly known as a Classical Nova outbursts.
Within
hours after outburst, the white dwarf is engulfed in an envelope of optically thick material that is driven away from the white dwarf by radiation pressure. Early in the evolution, the nova is bright in optical, but the peak of the spectral energy distribution shifts to higher energies as the mass ejection rate (and thus the opacity) decreases, exposing successively hotter layers (see, e.g., \citealt{bode2008}). A few weeks after outburst, the nova becomes bright in X-rays, and the observed X-ray spectra resemble those of the class of Super Soft Binary X-ray Sources (SSSs).

The nova V4743\,Sgr was discovered in September 2002 by \cite{haseda2002}. The first measured envelope ejection velocities exceeded 1200\,km\,s$^{-1}$ \cite{kato2002}. \cite{nielbock2003} studied the nova at $\lambda$1.2\,mm with the SEST instrument and concluded that the dominant emission source is the heated dust rather than free-free emission. The spectrum of the Nova was classified as Fe II type by \cite{morgan2003}. \cite{petz2005} found significant absorption from Fe and N from atmosphere modelling to a SSS X-ray spectrum taken with \chandra\ 180 days after outburst, that had been presented earlier by \cite{ness2003}. Ness et al. found large-amplitude fast variability with a period of $\sim 22$ minutes with clear overtones of this signal in the periodogram. During this observation a strong decline in X-ray brightness was observed with simultaneous spectral change from continuum spectrum to emission lines. The count rate dropped from 44 counts per second to 0.6 within $\sim 6$\,ks and stayed low for the rest of the observation (Ness 2010, in preparation). \cite{kang2006} presented CCD unfiltered optical photometry analyses from observations taken in 2003 and 2005. They detected a period of 6.7\,h which was interpreted as the orbital period of the underlying binary system. Six observations taken in short succession between days 1003 and 1011 days after outburst showed fast variability with a period of $\sim 24$ minutes. The authors attributed this signal to the beat period between the orbital and spin period of the white dwarf, where the 22 minute signal present in X-ray (\citealt{ness2003}) was assumed to be the spin period of the central white dwarf. Sophisticated X-ray period studies were presented by \cite{leibowitz2006}. They reanalysed the \chandra\ data taken on day 180, together with an \xmm\ observation taken on day 196. The large-amplitude variations were also found in the \xmm\ data set.
They detected at least 6 frequencies on day 180 and at least 12 on day 196.
Most of the variability was explained by a combination of oscillations at a set of discrete frequencies, and the main feature with a period of $\sim 22$ minutes has a double peak in the periodogram. At least 5 signals were constant between the two observations. \cite{leibowitz2006} suggested that the main feature in the periodogram at $\sim 22$ minutes ($\sim 0.75$\,mHz) is related to the white dwarf spin and that the other observed frequencies are produced by non-radial white dwarf pulsations.

In this paper we present timing analyses of all \chandra\ and \xmm\ observations, including those already analysed, for consistency. The main goal is to characterise the main feature at $\sim 0.75$\,mHz and its possible multiple structure as well as following the evolution of the oscillations. We address the question whether these variations are produced by the white dwarf rotation or by its pulsations. In Sect.~\ref{obs} we present all analysed X-ray data sets, and in Sect.~\ref{period_anal_tech} we describe our period analysis methods. The results are described in Sect.~\ref{results}. This section is structured
into the presentation of two different methods described in
Sect.~\ref{period_anal_tech}, the evolution of the oscillation amplitude of
the main signal, studies on beat periods, and other signals within a larger
frequency radius around the main signal as well as overtones.
Our results are discussed in Sect.~\ref{discussion}, and a summary with
conclusions is given in Sect.~\ref{summary}.

\section{Observations}
\label{obs}

In Table~\ref{obs_tab} we sumarise the analysed observations of the nova
V4743\,Sgr. The \chandra/ACIS (\citealt{nousek1987}) light curve for day 50.2
was extracted from the level-two events file using a circular extraction
region with radius 20 pixels. The \chandra/LETGS (\citealt{brinkman2000})
light curves for days 180, 302, 371 and 526 were extracted from the
non-dispersed photons (0th order), and the \xmm/RGS light curve
for day 196 was extracted from the dispersed photons. The two
\xmm/MOS (\citealt{turner2001}) light curves for days 742 and
1286 were extracted from the events file using an extraction region with
radius 200 pixels. In both observations, the source was on a chip gap in
the pn, and only the MOS1 observations can be used.

In the \xmm\ observation on day 742, additional optical/UV data from
the Optical Monitor (OM, \citealt{talavera2009}) are available. Five light
curves have been taken in 
series using the filters U, B, UVW1, UVM2, and UVW2 at
50-second time resolution. The different sensitivities of each filter
are not of relevance to this paper, and the light curves shown in
Fig.~\ref{optical} are normalised. For the other \xmm\ observations,
the source was either too bright, or no timing mode was used.

\begin{figure}
\includegraphics[width=84mm]{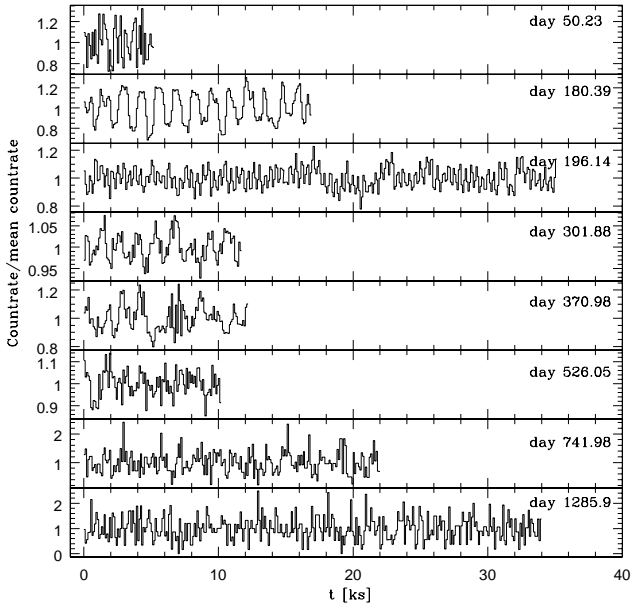}
\caption{Detrended and normalised X-ray light curves of all observations, indicated
by the days after outburst given in the right legend of each panel (for
observation log see Table~\ref{obs_tab}). The detrending and normalisation was
done using a 5-th order polynomial.}
\label{obs_fig}
\end{figure}

\begin{figure}
\includegraphics[width=84mm]{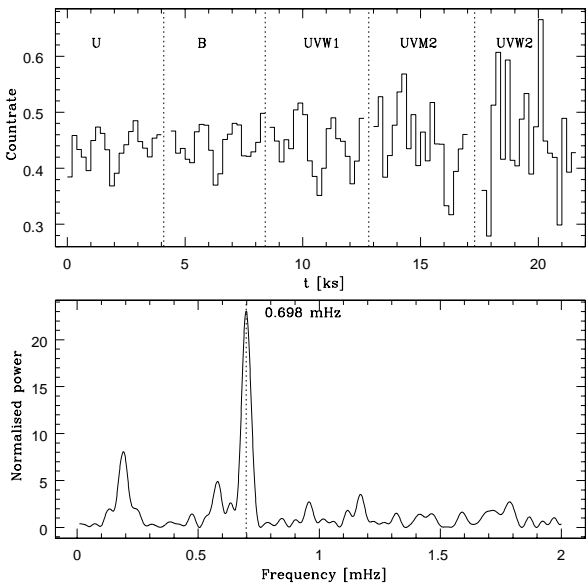}
\caption{Optical/UV light curve taken in series with the \xmm/OM on day
742 using different filters (upper panel) and corresponding Lomb-Scargle
power spectrum with candidate frequency marked by dotted line (lower panel).}
\label{optical}
\end{figure}

Our focus is on the $\sim 0.75$\,mHz period, and we thus removed all long-term variations by dividing the light curves by fifth-order polynomial fits. In Fig.~\ref{obs_fig} we show the de-trended, rebinned (100-s bins) light curves. For day 180, we only analyse the first 17 ks of data in order to avoid contamination by the steep decline that occurred shortly after (\citealt{ness2003}).
No such event occurred in any of the other observations, and the full data
sets were used for all other observations.

For consistency checks, we have also extracted separate light curves
for a hard and a soft band (15-30\,\AA\ and 30-40\,\AA, respectively),
filtering on the wavelengths in the level 2 event files.

\begin{table}
\caption{Journal of observations. Given are 'Day', days after outburst,
'Mission', name of observing mission, '$t$', exposure time in seconds, '$N$', the number of time bins, '$\Delta t$', the duration of one time bin in seconds, and 'cr', and mean count rate (counts per second). For day 180 we excluded the declining part of the light curve. In the text we refer to observations using days after outburst. For the optical/UV observation on day 742 we do not quote the mean count rate because of different filters used in the observation.}
\begin{center}
\begin{tabular}{llrrrl}
\hline
\hline
Day & Mission & $t$ & $N$ & $\Delta t$ & cr\\
\hline
50.23 & \chandra & 5250 & 211 & 25 & $0.2815\,\pm\,0.0001$\\
180.39 & \chandra & 17000 & 681 & 25 & $35.2495\,\pm\,0.6472$\\
196.14 & {\it XMM}/RGS & 35280 & 1408 & 25 & $52.0715\,\pm\,0.4411$\\
301.88 & \chandra & 11800 & 473 & 25 & $35.0487\,\pm\,0.1449$\\
370.98 & \chandra & 12300 & 493 & 25 & $17.7251\,\pm\,0.2687$\\
526.05 & \chandra & 10275 & 412 & 25 & $3.7795\,\pm\,0.0345$\\
741.98 & {\it XMM}/MOS1 & 22100 & 885 & 25 & $0.0875\,\pm\,0.0016$\\
1285.9 & {\it XMM}/MOS1 & 34050 & 682 & 50 & $0.0579\,\pm\,0.0016$\\
\hline
741.98 & {\it XMM}/OM & 21722 & 206 & 100 & --\\
\hline
\end{tabular}
\end{center}
\label{obs_tab}
\end{table}

\section{Period analysis techniques}
\label{period_anal_tech}

The goal of this work is to study the dominant 0.75 mHz feature detected
by \cite{ness2003} and analysed in more detail by \cite{leibowitz2006}. The principal problem is the multifrequency nature of this
dominant peak on day 196 (\citealt{leibowitz2006}). While \cite{ness2003}
reported only one period in the power spectrum for day 180,
\cite{leibowitz2006} found two nearby frequencies, using a least square
fitting approach. We
note that the observation on day 180 is much shorter than that taken on
day 196 (see Fig.~\ref{obs_fig} or Table~\ref{obs_tab}), and a clear
separation of multi-frequency oscillations is more difficult.

For our period analysis we use three different methods. First, we apply the standard Lomb-Scargle algorithm (\citealt{scargle1982}). This method calculates the normalised power for every frequency from the interval of interest with a selected frequency step. The result is a 2-D periodogram showing normalised power vs. tested frequency. The significant maxima are the significant signals.

We also use a linear combination of one or multiple sine curves that is
fitted to the light curve. If $t_i$ defines a time grid, with $i$
indicating the index of each time bin, the model $s_i$ is defined on
the same grid as
\begin{equation}
s_i =  P_i + \overline{B_i} + \sum^n_{j=1} A_j {\rm sin} (2 \pi t_i f_j + \phi_j)\,,
\label{sincurve}
\end{equation}
where $P_i$ denotes a 5-th order polynomial used for detrending,
$\overline{B_i}$ is the mean value of the instrumental background counts
in each time bin, and $n$ is the number of assumed sine curves with $A_j$,
$f_j$, and $\phi_j$ being the respective amplitudes, frequencies, and phase
shifts.

Instead of standard least square fitting (i.e., minimisation of $\chi^2$),
we use maximum likelihood iteration. The reason is that some of our light
curves have low count rates for which Poissonian noise has to be assumed.
With $s_i$ given in Eq.~\ref{sincurve} and $c_i$ the observed light curve
(in units counts per bin on the same time grid), the likelihood
is defined as
\begin{equation}
L =-2 \sum_{i=1}^N [c_i {\rm ln} (s_i) - s_i]\,.
\label{like}
\end{equation}
Note that $c_i$ is not background subtracted, because that would destroy
the Poissonian nature of the data. Instead, we do forward fitting, i.e.,
instead of subtracting the background $\overline{B_i}$, we add it to
the model as defined in Eq.~\ref{sincurve}.

In the fitting procedure, the parameters $A_j$, $f_j$, and $\phi_j$ are
iterated to minimise $L$ as defined in Eq.~\ref{like}. The 1-$\sigma$,
2-$\sigma$, and 3-$\sigma$ uncertainties (equivalent to 68.3\%,
95.4\% and 99.7\% confidence
levels) for the frequencies $f_j$ can be determined by interpolating the
$L(f_j)$-surface to values of $L+\Delta L_1$, $L+\Delta L_2$, and
$L+\Delta L_3$, respectively, where the values of $\Delta L_{1,2,3}$
depend on the number of degrees of freedom (thus the number of sine
curves, $n$). For $n=1$, $\Delta L_{1,2,3}$ assumes values of
1.0, 4.0, and 8.85, respectively, while for $n=2$, $\Delta L_{1,2,3}$
takes the respective values of 2.30, 6.16, and 11.6.

The uncertainties in the Lomb-Scargle method have been derived from the
false alarm probability (\citealt{horne1986}) of 0.3\%, which is
equivalent to the 3-$\sigma$ confidence. In cases where the errors are
not constrained within 3-$\sigma$ confidence because of low normalised
power of the peaks, we used higher false alarm probabilities of 4.6\%
and 31.7\%, equivalent to 2 and 3-$\sigma$ confidence, respectively.
In the cases with even lower normalised power we quote only the nominal
values without errors.

\begin{figure}
\includegraphics[width=55mm,angle=-90]{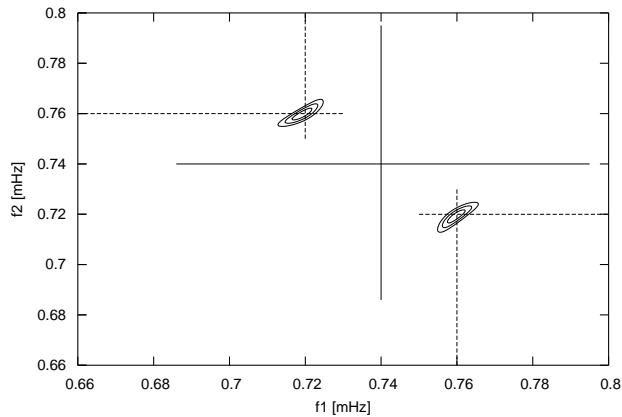}
\caption{Contour plots of the 2-D sine fit period analysis applied to a test light curve. The test light curve is modulated with the two frequencies 0.72 and 0.76\,mHz (marked by dotted lines), and is sampled as the day 180 light curve. The lines are the contours of the likelihood from sine fitting of the analysed data set and the axes are the tested intervals of the frequencies $f_1$ and $f_2$. Shown are the 1-$\sigma$, 2-$\sigma$ and 3-$\sigma$ contours.
The solid lines intersecting at 0.740\,mHz indicates the result (with errors equivalent to 0.3\% false alarm probability) from the Lomb-Scargle algorithm.
}
\label{power_run2_test2}
\end{figure}

\begin{figure}
\includegraphics[width=55mm,angle=-90]{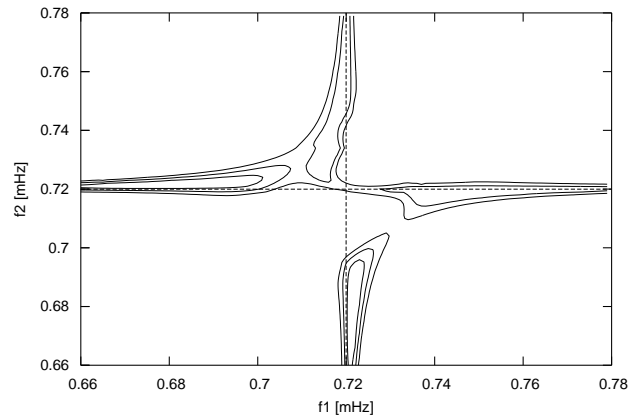}
\caption{Same as Fig.~\ref{power_run2_test2} for a test light curve modulated with a single frequency of 0.72\,mHz (marked by dotted lines) and sampled as the day 180 light curve.}
\label{power_run2_test1}
\end{figure}

In this paper we refer to the $n=1$ case as the '1-D method',
while we call the $n=2$ case the '2-D method'. We illustrate the
results of our 2-D method in the form of contour plots. Examples of
such contour plots are shown in Figs.~\ref{power_run2_test2}
and \ref{power_run2_test1} for two different synthetic light curves
which are sampled as the \chandra\ light curve taken on day 180
with comparable amplitude. We modulated the synthetic light curves
with one and two frequencies and added Poissonian noise and then
applied the 2-D method to each synthetic light curve. In
Fig.~\ref{power_run2_test2}, the result for the synthetic light curve
that has been modulated with two frequencies is shown. The input
frequencies are $f_1=0.72$ and $f_2=0.76$\,mHz (see horizontal and
vertical dotted lines in Fig.~\ref{power_run2_test2}), and it can be seen
that these values can be recovered. Closed contour lines, centred around
the coordinates ($f_1$, $f_2$) and ($f_2$, $f_1$) are the result.
In contrast the case of a synthetic light curve that is modulated with
only one frequency, yields a different contour plot (see
Fig.~\ref{power_run2_test1}). Again, the input frequency of
0.72\,mHz is marked by dotted lines, and open contours around these
lines are the result.

\begin{figure}
\includegraphics[width=84mm]{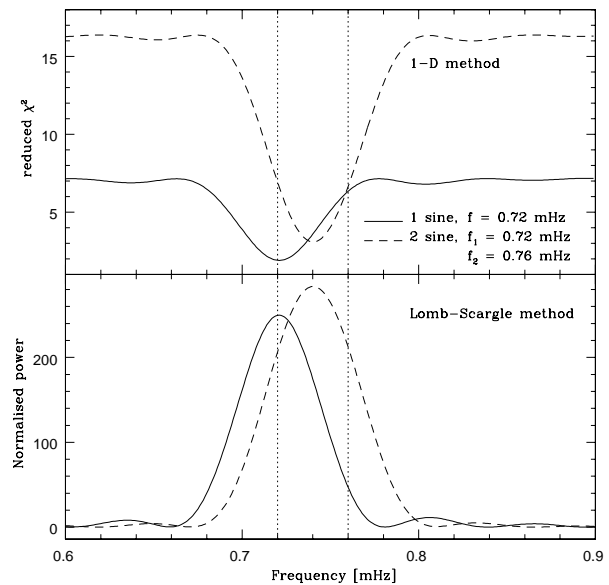}
\caption{$\chi^2$ and Lomb-Scargle normalised power curves for a range of frequencies around the input
frequencies of two synthetic light curves. The solid lines indicate
the $\chi^2$ curves for the synthetic light curve that was modulated
with a single frequency at 0.72\,mHz, while the dashed curves indicate
the $\chi^2$ curves for the  synthetic light curve that was modulated
with two frequencies at 0.72\,mHz and 0.76\,mHz (see lower right
legend in top panel). In the top panel, the respective $\chi^2$ curves
for the 1-D method are shown, and in the bottom panel the same for
the Lomb-Scargle results. The single frequency can be recovered with
both methods, but the 2-frequency light curve yields only a single
frequency in between the two input frequencies.
}
\label{1-D_tests}
\end{figure}

Meanwhile, both, the 1-D and the Lomb-Scargle methods, applied to the
light curve that is modulated with two frequencies, yield only
a single period at a frequency of 0.740\,mHz (Fig.~\ref{1-D_tests}).
In Fig.~\ref{power_run2_test2}, this frequency is marked with the
horizontal and vertical lines, the length of which indicate the
error bars. This demonstrates that
our 2-D method allows us to discriminate between light curves that are
modulated with one and with two frequencies.

\section{Results}
\label{results}

 We have employed the techniques described in Sect.~\ref{period_anal_tech}
to all data sets and summarise our results in Table~\ref{det_signals}. In
the following subsections we describe the results from the 1-D and 2-D
methods. We focus on the multifrequency nature of the dominant peak at
0.75\,mHz employing the 1-D method and the 2-D method in Sects.~\ref{1dmethod}
and \ref{2dmethod}, respectively. We discuss the role of the beat period
in relation to the exposure time in Sect.~\ref{sect:beat}. The evolution
of the oscillation amplitudes is determined in Sect.~\ref{sect:ampl}, and
the expected harmonic frequencies of 0.75\,mHz in Sect.~\ref{othersignals}.

\subsection{1-D methods}
\label{1dmethod}

\begin{figure}
\includegraphics[width=84mm]{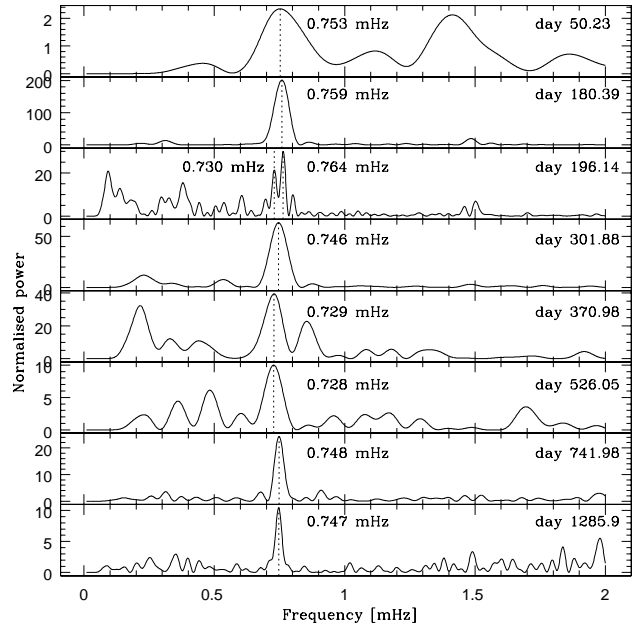}
\caption{Power spectra of all X-ray data sets using the Lomb-Scargle algorithm.
The candidate frequencies summarised in Table~\ref{det_signals} are marked by dotted lines in each panel.}
\label{LSpower_all_run}
\end{figure}

The Lomb-Scargle and 1-D sine fitting methods give similar results.
In Fig.~\ref{LSpower_all_run} we show the Lomb-Scargle periodograms
of all light curves. Except for the observation taken on day 196, all
observations yield only a single peak around 0.75\,mHz. For day 196, a more
complicated structure can be seen.
"Is such complex
nature of the main peak present only at this day?" Such a question
motivates this project of searching for the two central
dominant peaks in the other light curves. In the L-S column of
Table~\ref{det_signals} we give the signals detected with the
Lomb-Scargle method near the value of 0.75\,mHz for all light curves.

Following the tests with the synthetic light curves described in the
previous section, the non-detection of a multifrequency structure with
the Lomb-Scargle analysis does not necessarily rule out the presence of
two or more nearby frequencies. In order to test for the presence of
such additional signals, we have
pre-whitened the data with the detected frequencies. By subsequent period search we find for day 180
another dominant peak at ($0.701\,\pm\,0.033$)\,mHz
with the error determined from the 0.3\% false alarm probability. For day 196, the subtracted
peak disappeared from the periodogram while the second peak remained with
practically the same normalised power. For the other observations
the main signal completely disappeared
after the pre-whitening, hence no multifrequency structure can be identified.

\begin{figure}
\resizebox{\hsize}{!}{\rotatebox{270}{\includegraphics{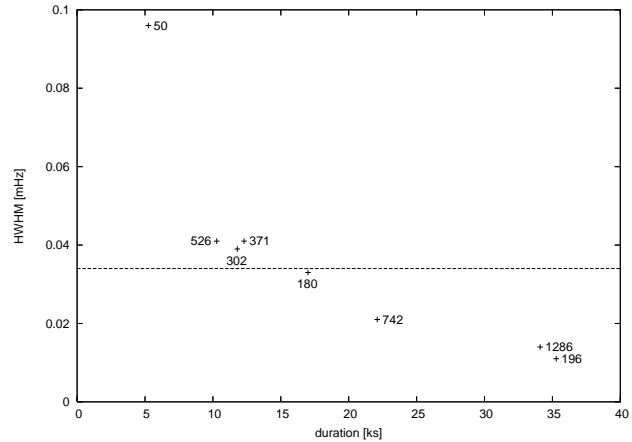}}}
\caption{Half width at half maximum of the peaks in the Lomb-Scargle
power spectrum of X-ray data versus respective exposures times. Long exposure times
yield narrow peaks, and multiple signals can only be detected if
the exposure time covers at least one beating cycle.
The horizontal line represents the difference
between the detected frequencies from day 196.
}
\label{LSwidth}
\end{figure}

In Fig.~\ref{LSwidth} we show the half width at half maximum for the peaks in the
Lomb-Scargle power spectra as resolution indicator versus the respective data durations. The difference
between the detected frequencies
from day 196 is shown by the horizontal dotted line. The peaks in the Lomb-Scargle
power spectra are broader than the day 196 frequency
for days 50, 302, 371, and 526. Thus, if there were two frequencies as
during day 196, these would not be detectable because
the peaks are too broad to resolve them. The width for day 180 is marginally
close to the frequency difference for day 196, but this is apparently not
sufficient to resolve two frequencies. However, for days 742 and 1286, the
peaks are significantly narrower, and two close signals could be
detected if their difference was of the same order as for day 196.
We can thus confidently conclude from the 1-D methods that only a single
signal was present during the last two observations.

The Lomb-Scargle power spectrum of the \xmm/OM light curve taken on
day 742 is shown in Fig.~\ref{optical}. The frequency found is
$0.698 \pm 0.020$\,mHz. Pre-whitening of the data with this frequency yields
a complete disappearance of any other signal. Therefore, this is the
only signal.

\begin{figure}
\includegraphics[width=84mm]{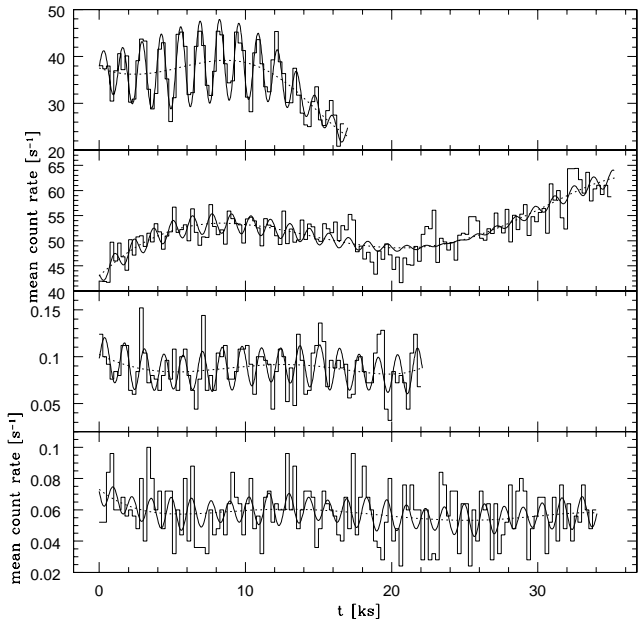}
\caption{Original X-ray light curves and best-fit 2-D models for days 180, 196,
742, and 1286. For illustration purposes, the data and model are rebinned
to a time bin size of 250\,sec.
The solid line is the 2-D sine best fit and the dotted line is the detrending polynomial.}
\label{fits}
\end{figure}

\subsection{2-D method}
\label{2dmethod}

 In contrast to the Lomb-Scargle method, the 2-D sine fitting method is
a direct fitting approach, and the fits can be illustrated. In Fig.~\ref{fits},
we show the original (not detrended) light curves and the best-fit models,
$s_i$, according to Eq.~\ref{sincurve}, for the observations taken on days
180, 196, 742, and 1286. In each panel, the dashed line indicates the 5-th
order polynomial, $P_i$, that was used
for detrending. The light curve taken on day 196 is the most complex of all,
and \cite{leibowitz2006} reported the presence of multiple
frequencies from the power spectrum and from least squares fitting.
They also reported that they resolved two frequencies (0.728 and
0.782\,mHz) with least squares fitting to the day-180 data, yielding
better fits than single- or three-frequency fits.

 In order to investigate the detailed structure of the 0.75-mHz signal,
we closely inspect our contour plots for each fit. As a detection criterion
we choose the 99.7\% confidence, i.e., the 3-$\sigma$ significance level.
In Table~\ref{det_signals} we list the resulting frequencies with
corresponding 3-$\sigma$ uncertainties.

 The first observation on day 50 is too short to yield any significant
detection of a second period. Even the presence of the 0.75-mHz signal
itself
as a single frequency can only be established on the 1-$\sigma$ level,
and the uncertainties given in Table~\ref{det_signals} for this
observation are only 1-$\sigma$ errors.

\begin{figure}
\includegraphics[width=55mm,angle=-90]{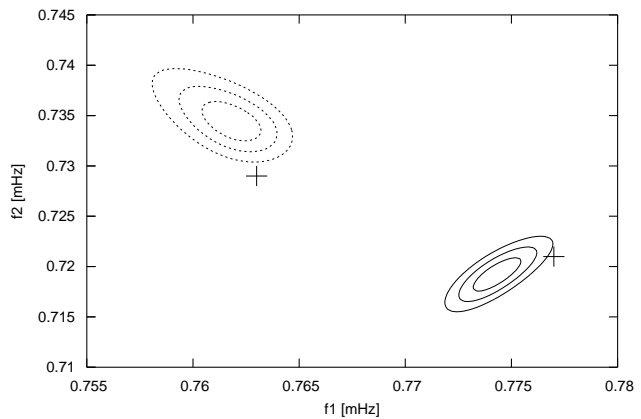}
\caption{Same as Fig.~\ref{power_run2_test1} for days 180 (solid line) and 196 (dashed line). The values derived by Leibowitz et al. (2006) are marked by two crosses.}
\label{power_run2_run3_detail}
\end{figure}

 The result of the analysis of the
day-180 light curve is shown in Fig.~\ref{power_run2_run3_detail} (solid
contour lines in the bottom right). We find closed contours around the
pair of frequencies $0.719\,\pm\,0.004$ and $0.774\,\pm\,0.003$\,mHz.
This is consistent within the errors with our result from the
pre-whitening
(Sect.~\ref{1dmethod}). We emphasise, however,
that the error from the 1-D method is much larger,
i.e. 0.062\,mHz (Tab.~\ref{det_signals}).
The result from
\cite{leibowitz2006} is marked with a plus symbol in
Fig.~\ref{power_run2_run3_detail}. While it is just outside of our
3-$\sigma$ contour lines, it is yet close enough to argue that the
two independent results are reasonably consistent
within the combined errors (Leibowitz et al. have given no errors,
which we estimate from their periodogram as $\sim 0.02$\,mHz).

For day 196, two peaks are clearly detected, and the result from
our 2-D approach is included in the top left corner of
Fig.~\ref{power_run2_run3_detail} (dotted contour lines).
The two frequencies are well localised and form a deep minimum at
$0.734\,\pm\,0.005$ and $0.762\,\pm\,0.004$\,mHz. These values also
agree with Leibowitz et al. (2006) within the combined uncertainties.
The contour lines of day 180 (solid lines) and from day 196 (dotted
lines) do not overlap, and the frequencies are thus significantly
different. Apparently, the two detected frequencies are closer
together on day 196.

 For the well-exposed observations taken on days 180 and 196,
we have also analysed separate light curves extracted from a soft
band (30-40\,\AA) and a hard band (15-30\,\AA). The soft light curve
for day 180 delivers exactly the same result. The frequencies
derived from the other three light curves are consistent within
the statistical 3-$\sigma$ uncertainty ranges, and the conclusion
that the two signals on days 180 and 196 are different is robust
against the chosen energy band.

\begin{figure}
\includegraphics[width=55mm,angle=-90]{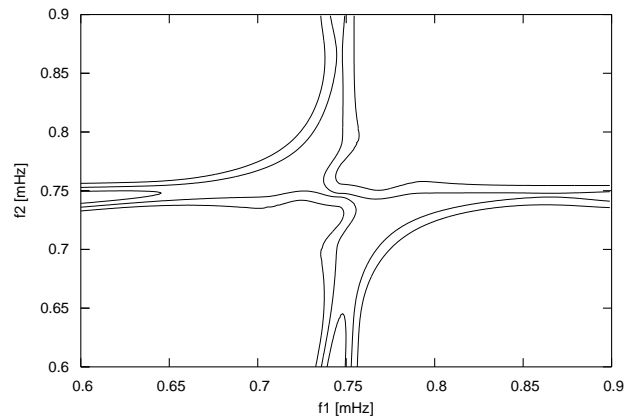}
\caption{Same as Fig.~\ref{power_run2_test1} for day 302. The shape contour lines
resembles that in Fig.~\ref{power_run2_test1}, indicating that only a single
frequency is present with $f=0.74\,\pm\,0.01$\,mHz
(see Table~\ref{det_signals}).}
\label{power_run4}
\end{figure}

 For the observations taken on days 302, 526, 742, and 1286, no indication for
a double frequency structure can be detected. This is illustrated in
Fig.~\ref{power_run4} and Figs.~\ref{power_run6}--\ref{power_run8},
respectively, where only a single frequency is found with the 2-D
method. While for days 526, 742 and 1286, closed contour lines are
encountered at the 1-$\sigma$ significance level, this can not be
considered significant enough for a secure detection of two frequencies.
However, the detection of a single frequency is significant at the
99.7\% confidence level.

\begin{figure}
\includegraphics[width=55mm,angle=-90]{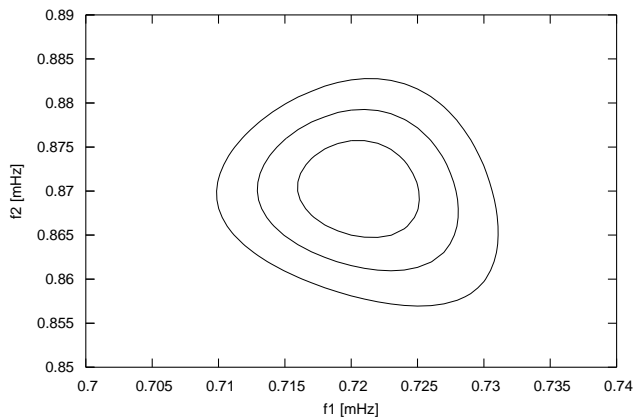}
\caption{Same as Fig.~\ref{power_run2_test1} for day 371. The
closed contour lines indicate two frequencies at $f_1=0.72$\,mHz and
$f_2=0.87$\,mHz. The latter is not near the main 0.75-mHz signal and is
thus beyond the scope of this paper.}
\label{power_run5_detail}
\end{figure}

\begin{figure}
\includegraphics[width=55mm,angle=-90]{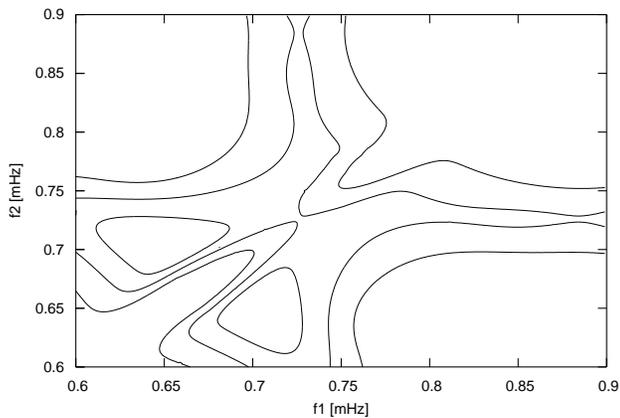}
\caption{Same as Fig.~\ref{power_run2_test1} for day 526.}
\label{power_run6}
\end{figure}

\begin{figure}
\includegraphics[width=55mm,angle=-90]{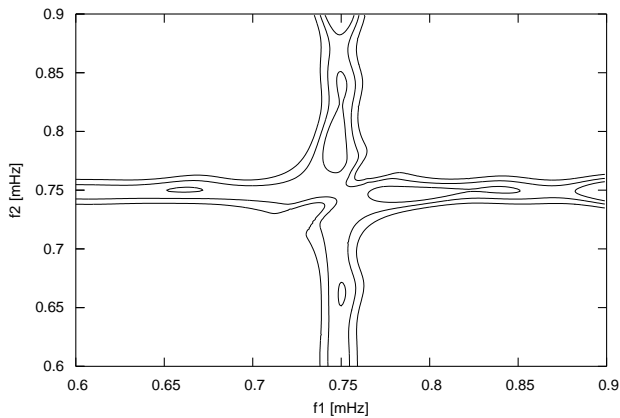}
\caption{Same as Fig.~\ref{power_run2_test1} for day 742 in X-ray band.}
\label{power_run7}
\end{figure}

\begin{figure}
\includegraphics[width=55mm,angle=-90]{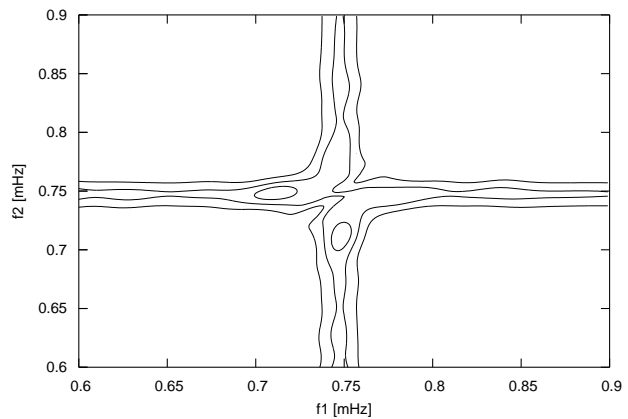}
\caption{Same as Fig.~\ref{power_run2_test1} for day 1286.}
\label{power_run8}
\end{figure}

Meanwhile, for day 371 a deep minimum with closed contours is found
(Fig.~\ref{power_run5_detail}), indicating the presence of two frequencies
($0.721\,\pm\,0.010$)\,mHz and ($0.870\,\pm\,0.011$)\,mHz.
Both signals are also visible in the Lomb-Scargle power spectrum.
The 0.87-mHz frequency is in fact unrelated to the main signal and is
also present in the other datasets (see more detailed discussion in
Sect.~\ref{othersignals} and Table~\ref{tab:othersigs}).
The quality of the data does not allow us to resolve the 0.721\,mHz
feature
into two individual signals. Within the measurement precision,
we thus only have a single frequency of 0.721\,mHz for day 371.

\begin{figure}
\resizebox{\hsize}{!}{\includegraphics{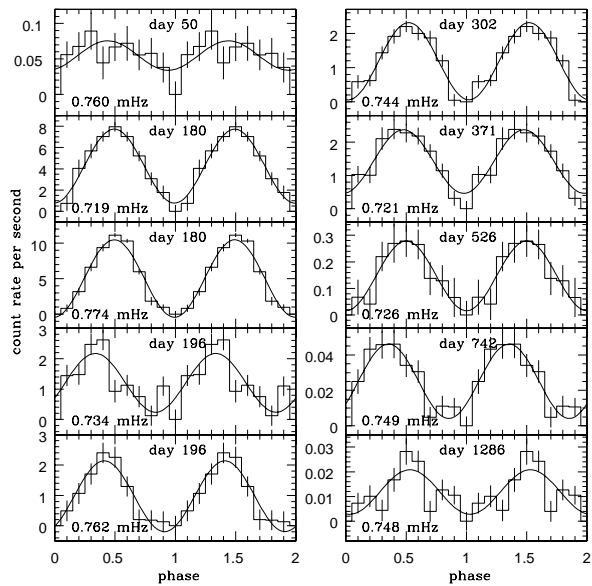}}
\caption{Phased light curves (detrended, binned into 50-s bins)
for the frequencies found by 2-D sine fitting (see 3-rd column of
Table~\ref{det_signals}).
}
\label{phase}
\end{figure}

 In Fig.~\ref{phase} we show the phased light curves, assuming the
resultant frequencies reported in this section. The shape of the
phased light curves are clearly sinusoidal in all cases, and our
approach of sine curves is thus justified. One can also see from
this figure that the frequencies are all well sampled.
\begin{table}
\caption{Detected X-ray signals around the 0.75\,mHz frequency, between
0.72 and 0.78\,mHz (Other signals are listed in
Table~\ref{tab:othersigs}).
The 'L-S' column lists the results from the Lomb-Scargle analysis
with errors computed from the half width at half maximum in the power
spectrum. The '2-D sine fit' column lists the results from our 2-D sine
fitting analysis with errors calculated from the 3-$\sigma$ contours.
In the last column the results of Leibowitz et al (2006) are
give for comparison.}
\begin{center}
\begin{tabular}{lcccr}
\hline
\hline
Day& L-S & 2-D sine fit &  Leibowitz\\
&mHz&mHz&et al. (2006)\\
\hline
50 & $(0.753)^b$ & $0.760\,\pm\,(>0.053)^a$ & -- \\
\hline
180 & -- & $0.719\,\pm\,0.004$ & 0.721 \\
 & $0.759\,\pm\,0.062$ & -- & -- \\
 & -- & $0.774\,\pm\,0.003$ & 0.777 \\
\hline
196 & $0.730\,\pm\,0.009$ & $0.734\,\pm\,0.005$ & 0.729 \\
 & $0.764\,\pm\,0.011$ & $0.762\,\pm\,0.004$ & 0.763 \\
\hline
302 & $0.746\,\pm\,0.113$ & $0.744\,\pm\,0.012$ & -- \\
\hline
371 & $0.729\,\pm\,0.051$ & $0.721\,\pm\,0.011$ & -- \\
\hline
526 & $0.728\,\pm\,(>0.020)^a$ & $0.726\,\pm\,0.028$ & -- \\
\hline
742 & $0.748\,\pm\,0.019$ & $0.749\,\pm\,0.011$ & -- \\
\hline
1286 & $0.747\,\pm\,(>0.008)^a$ & $0.748\,\pm\,0.011$ & -- \\
\hline
\end{tabular}
\end{center}
$^a$unconstrained on 3-$\sigma$ level and only 1-$\sigma$ level given\\
$^b$unconstrained on 1-$\sigma$ level and only the nominal value is given
\label{det_signals}
\end{table}

\begin{figure}
\includegraphics[width=55mm,angle=-90]{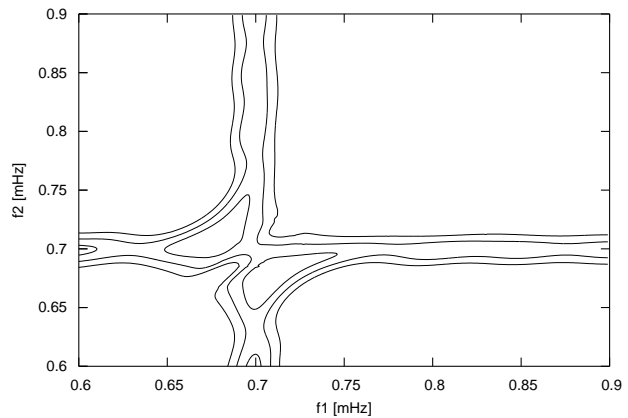}
\caption{Same as Fig.~\ref{power_run2_test1} for optical/UV from \xmm/OM data obtained on day 742.}
\label{power_optical}
\end{figure}

Finally, the analysis of our optical/UV data taken on day 742 is shown on Fig.~\ref{power_optical}. Only one frequency is found with value $0.699 \pm 0.015$\,mHz which is consistent with our Lomb-Scargle finding.

\subsection{Beat Periods}
\label{sect:beat}

\begin{figure} 
\resizebox{\hsize}{!}{\includegraphics{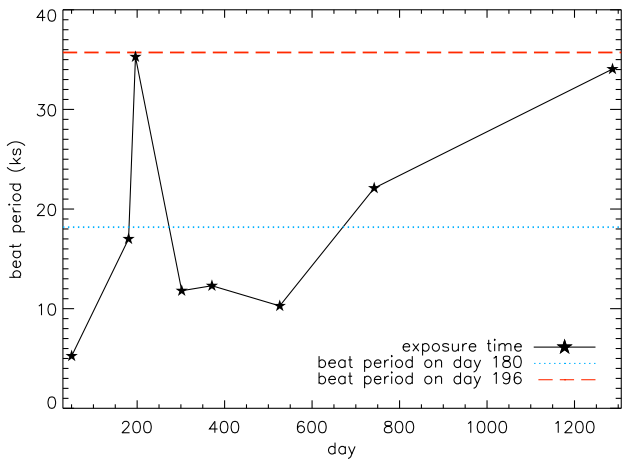}}
\caption{Illustration of detectability of two signals in our observations.
The star symbols, connected by the black line, are the exposure times
listed in Table~\ref{obs_tab}. If the beat period between two signals
is longer than the exposure time, then these frequencies can not be
detected as two separate signals, because the exposure time does not
cover at least one beat cycle.
The blue dotted and red dashed lines indicate the beat periods for days
180 and 196. The limit imposed by too short exposure time prohibits the
detection of these two frequencies on days 50, 301, 371, and 526. But
for days 742 and 1286, at least the two frequencies found for day 180
would be detectable, unless the ratio of the amplitudes of the two
signals is less than 1:2 for a 99.7\% confidence detection
(for a 95,4\% and 90\% confidence detection, an amplitude ratio of
1:3 and 1:4 is sufficient, respectively).
}
\label{beat}
\end{figure}

The superposition of two sine functions with different frequencies $f_1$
and $f_2$ results in beating at frequency $|f_1 - f_2|$. We expect that
two frequencies can only be separated with a given observation duration
if it is longer than the beat period, $1/|f_1 - f_2|$. The only observations
for which we are able to find two frequencies are those on days 180 and
196. The difference between these two frequencies are 0.055\,mHz and
0.028\,mHz, respectively (see Table~\ref{periods}), and the expected
beat periods are
($18.2\,\pm\,0.1$)\,ks and ($35.7\,\pm\,0.2$)\,ks, respectively. As can be
seen from Table~\ref{obs_tab}, the exposure times of these two observations
are sufficiently long to cover one cycle of the beat period. However,
the observations taken on days 301, 372, and 526 are significantly
shorter, and two frequencies can not easily be separated.
We tested this statement by analysing a synthetic light curve, sampled and
noised as day 302 and modulated with two close frequencies of 0.72 and 0.76
mHz. Our 2-D method did not find a closed minimum within 3-$\sigma$ lines and
the global minimum in the goodness contours is found for the combination of
values of 0.678 and 0.745 mHz. Since the exposure
times for days 371 and 526 are of the same order with lower signal to noise,
we conclude that the exposure times for days 302, 371, and 526 are too
short and the data are too noisy for a positive detection of two signals.

 The importance of the exposure time for the resolution of two frequencies
is illustrated in Fig.~\ref{beat}, where we plot the exposure
times (in units ks) with the star symbols, connected by the black line and
the beat periods for days 180 and 196 for comparison.
The best resolving power can be achieved with the
exposure times on days 196 and 1286 when a full beating cycle is covered.
Two signals with a frequency difference as observed for days 180 or 196
can not be detected in any of the observations taken on days 50, 301, 371,
and 526.
However, the observation taken on day 742 was long enough to at least
recover two frequencies if $|f_2-f_1|$ was as large as on day 180. For
day 1286, the situation detected for day 196 would have been
detectable. This conclusion is valid with comparable modulation amplitudes,
but different amplitude ratios can affect the detectability. To test the
effect, we have produced a synthetic data set sampled as day 742 and 1286,
with amplitudes taken from Table~\ref{amplitudes_tab} and with
Poissonian noise. With our 2-D method, we were able to identify both
signals at 0.72 and 0.76\,mHz for the amplitude ratios 1:2, 1:3 and 1:4
at a confidence level of 99.7\%, 95,4\%, and 90\%, respectively.

 If on days 301, 372, and 526 the main peak was split in two signals
as on days 180 or 196, then we would detect a value somewhere in between the
minimum and maximum value. We experimented with synthetic light
curves modulated by two signals of different amplitudes and found
that the resultant frequency would be closer to the signal
with the higher amplitude, and exactly in between if the amplitudes
were the same. Meanwhile, the formal 3-$\sigma$ measurement
uncertainties would be much smaller than the maximum frequency
difference between two signals that could be detected with
the given exposure times. As a consequence, the possibility
of having detected an average frequency between two signals
introduces a source of systematic uncertainty of the order of
half the inverse exposure time, because this value limits
the detectability of two signals.

 It is noteworthy from Fig.~\ref{beat} that the beat periods that
correspond to the two signals on days 180 and 196 are suspiciously
close to their respective exposure times.
This similarity might raise suspicions that the exposure
time could already predetermine the difference between the two signals.
In the case of such a bias the same two frequencies would have
resulted if the exposure times for days 180 and 196 had been the
same. While we can not test longer exposure, we have run several
tests using reduced exposure times for both light curves.
For day 196 we only found a single frequency around 0.76\,mHz if
only the first 17\,ks of the light curve are used. This can be
explained by the reduced coverage of the beat period, and
the detected two periods apparently require the long exposure
time. Our analyses of a sample of reduced data sets for day 180
always resulted in two detected frequencies. The resulting
values depend slightly on the selected subset of the light
curve, but no correlation between effective exposure time
and beat period is seen. The reason for the clear detection
of two signals for day 180 can be seen in the well-pronounced
sinusoidal form of the light curve with little contaminating
noise, as can be seen from Fig.~\ref{fits}. The Lomb-Scargle
method is limited by the requirement to cover at least one
beating cycle in order to detect the respective two close signals
(see Fig.~\ref{LSwidth}).
Meanwhile, our fitting approach is only limited by the quality
of the data, and two close signals can be detected, even if
only a fraction of the beating cycle is covered, if only the
data have sufficient signal to noise. For day 196, the count
rate is higher, but the light curve shows more irregular patterns,
and a clear detection of two periods requires a more
complete coverage of the beat period. From all these
considerations we conclude that the presence
of two frequencies is real in both cases and that
the similarity between
exposure time and beat period in the two cases must therefore
be a coincidence.

\subsection{Amplitudes}
\label{sect:ampl}

\begin{figure}
\includegraphics[width=84mm]{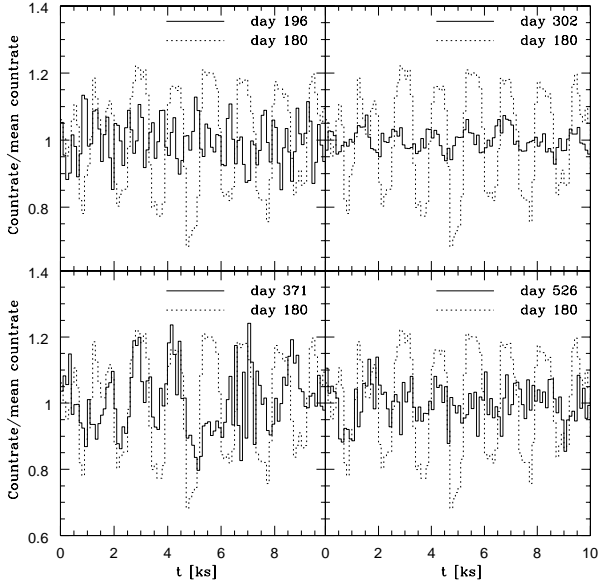}
\caption{Normalised X-ray light curves for days 196, 302, 371, and 526 as
indicated in the legends (solid lines) in comparison to day 180
(dotted line).}
\label{amplitudes_relat}
\end{figure}

 The comparison of normalised light curves in Fig.~\ref{amplitudes_relat}
illustrates that the oscillation amplitudes vary considerably between
the observations. While only parts of the light curves are shown for
better clarity, it is clear, e.g., from Figs.~\ref{obs_fig} and \ref{fits},
that the amplitude of the first 10 ksec is representative for the full
observation in each case. We include the normalised light curve for day
180 with a dotted line for reference. Clearly, on days 196 and 302, the
relative amplitude was much smaller than on day 180, and it seems that it
has almost recovered by day 372 but is smaller again on day 526.

 The amplitude parameters, $A_j$ in Eq.~\ref{sincurve}, are representative
of the evolution of the oscillation amplitude, as can be seen from, e.g.,
Figs.~\ref{fits} and \ref{phase}.

\begin{figure}
\includegraphics[width=84mm]{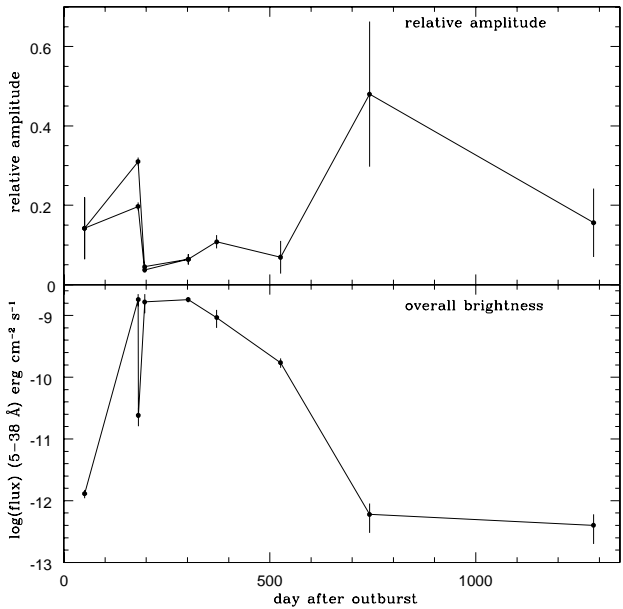}
\caption{Top: Time evolution of the relative amplitudes $A_j$ listed in the
last column of Table~\ref{amplitudes_tab}. Bottom: Evolution of X-ray fluxes
(see Sect.~\ref{spectra}) for comparison. The connection between
the decline on day 180 and a sudden drop in relative amplitude can be recognised.}
\label{aevol}
\end{figure}

 The evolution of the best-fit amplitude parameters $A_j$, relative to
the mean count rates given in the last column of Table~\ref{obs_tab},
is illustrated in Fig.~\ref{aevol}. The values are listed in the
last column of Table~\ref{amplitudes_tab}, where also the absolute
amplitudes are given for each observation and each frequency. The
absolute amplitudes have been determined directly from the detrended
light curves before normalisation. In the bottom panel of Fig.~\ref{aevol}
we show X-ray band fluxes (0.3-2.5\,keV) for comparison. The description
of flux measurements can be found in Sect.\ref{spectra}. The sharp drop
in brightness on day 180 first reported by \cite{ness2003} can be
identified in the bottom panel, and it coincides with a significant
reduction in oscillation amplitude. While the brightness has completely
recovered by day 196, the oscillation amplitudes remained low until the
end of the SSS phase after day 526.
\begin{table}
\caption{Amplitude parameters $A_j$ from Eq.~\ref{sincurve} for the days
given in the first column. Given are the X-ray absolute amplitudes (in units
counts per second, [cps]), and the X-ray amplitudes relative to the
mean count rates given in the last column of Table~\ref{obs_tab}, each
for the respective frequencies $f_j$ 
}
\begin{center}
\begin{tabular}{lccccr}
\hline
\hline
Day & $f$ [mHz] & absolute [cps] & relative \\
\hline
50 & 0.760 & $< 0.047$ & $< 0.168$\\
\hline
180 & 0.719 & $3.479 \pm 0.190$ & $0.099 \pm 0.005$\\
 & 0.774 & $5.465 \pm 0.190$ & $0.155 \pm 0.005$\\
\hline
196 & 0.734 & $0.967 \pm 0.162$ & $0.019 \pm 0.003$\\
 & 0.762 & $1.166 \pm 0.163$ & $0.022 \pm 0.003$\\
\hline
302 & 0.744 & $1.125 \pm 0.235$ & $0.032 \pm 0.007$\\
\hline
371 & 0.721 & $0.956 \pm 0.158$ & $0.054 \pm 0.009$\\
\hline
526 & 0.726 & $0.131 \pm 0.082$ & $0.035 \pm 0.022$\\
\hline
742 & 0.749 & $0.021 \pm 0.008$ & $0.237 \pm 0.095$\\
\hline
1286 & 0.748 & $0.009 \pm 0.005$ & $0.147 \pm 0.094$\\
\hline
\end{tabular}
\end{center}
\label{amplitudes_tab}
\end{table}

\subsection{Other signals}
\label{othersignals}

\begin{table}
\caption{Significant detections of other X-ray frequencies in the range
0.6-0.9\,mHz.
}
\begin{center}
\begin{tabular}{lccc}
\hline
\hline
day & 1st & 2nd & 3rd\\
\hline
180 &&&\\
\ $f\,[$mHz$]$ & $0.628 \pm 0.004$ & $0.669 \pm (0.006)^a$ & $0.863 \pm 0.004$\\
\ $A$(abs) & $0.777 \pm 0.196$ & $1.432 \pm 0.193$ & $1.991 \pm 0.193$\\
\ $A$(rel) & $0.022 \pm 0.006$ & $0.041 \pm 0.005$ & $0.056 \pm 0.005$ \\
196 &&&\\
\ $f\,[$mHz$]$ & $0.606 \pm 0.004$ & $0.703 \pm 0.007$ & $0.802 \pm 0.004$\\
\ $A$(abs) & $0.693 \pm 0.230$ & $0.520 \pm 0.232$ & $0.586 \pm 0.231$\\
\ $A$(rel) & $0.013 \pm 0.004$ & $0.010 \pm 0.004$ & $0.011 \pm 0.004$\\
371 &&&\\
\ $f\,[$mHz$]$ & -- & $0.671\pm 0.030$ & $0.870 \pm 0.013$\\
\ $A$(abs) & -- & $0.535 \pm 0.179$ & $0.692 \pm 0.157$\\
\ $A$(rel) & -- & $0.030 \pm 0.010$ & $0.039 \pm 0.009$\\ 
\hline
\end{tabular}
\end{center}
$^a$unconstrained on 3-$\sigma$ level and only 2-$\sigma$ level given
\label{tab:othersigs}
\end{table}

While the focus of this paper is the substructure and evolution of the
strongest signal around 0.75\,mHz, other nearby signals are present. 
\cite{leibowitz2006} found at least 12 significant frequencies for
day 196, and for five of these, similar frequencies were also found for
day 180. However, without uncertainty ranges it is difficult to decide
whether or not these frequencies are related to each other.

In Table~\ref{tab:othersigs} we list all 3-$\sigma$ detections
of frequencies in the range between 0.6 and 0.9\,mHz for days 180-371.
Basically, three significant frequencies can be detected in this range
for days 180 and 196. Their values differ on a statistically significant
level, and either these three frequencies are variable, or they occurred
at random and are not related to each other. However, the two frequencies
found for
day 371 are both in good agreement with the corresponding frequencies on day
180, and these frequencies may have undergone a similar evolution as
the main peak, i.e. reduction between days 180 and 196,
followed by a slow increase back to the level seen on day 180. This is
supported by the fact that the same trends of the third frequency in the
last column of Table~\ref{tab:othersigs} can be seen as in Fig.~\ref{pevol}.
This
behaviour could be related to the sharp decay that occurred on day 180.
The beat periods between the main
frequency of 0.75\,mHz and the signals at 0.67\,mHz and 0.87\,mHz
are both of order 10\,ks, which is short enough to
separate these frequencies from the main frequency, but too long in
order to be detectable in the power spectrum.
Along the definitions of amplitudes set out in Sect.~\ref{sect:ampl},
we calculated the relative and absolute amplitudes, which are also listed
in Table~\ref{tab:othersigs}.

Both, \cite{ness2003} and \cite{leibowitz2006} underlined the presence
of overtones at $\sim 1.5$\,mHz. We systematically searched for this signal
in all data sets with the same techniques as described in the previous sections,
and the results are presented in Table~\ref{det_overtones}. The results from
the Lomb-Scargle (L-S) and 2-D sine method agree well with each other.
The dataset taken on day 50 is again too short to yield useful results within
the 3-$\sigma$ contours, and the 1-$\sigma$ uncertainty range is given in
round brackets. For days 180 and 196, we clearly detected two frequencies
from both methods. The beat periods from the 2-D results are 17.8 and 19.6\,ks,
respectively, and such splitting of the main signal is thus detectable in the
given observing durations (Table~\ref{obs_tab}). We also found two
frequencies in the day 742 observation with both techniques, yielding a
beat period of $\sim 16$\,ks. With the 2-D method, both frequencies are only
significant on the 2-$\sigma$ level. For the
other light curves, only a single frequency, if any, could be identified.
\begin{table*}
\caption{First overtone X-ray frequencies expected around 1.5\,mHz.
For comparison, three linear combinations of the
observed ground harmonics from the second column in Table~\ref{det_signals}
are given in the last three columns.}
\begin{center}
\begin{tabular}{lccc|ccc}
\hline
\hline
&&&&\multicolumn{3}{|c}{Linear combinations of ground harmonics (Table~\ref{det_signals}):}\\
Day & L-S & 2-D sine fit & Leibowitz & $f_1+f_2$ & $2\times f_1$ & $2\times f_2$\\
&&&et al. (2006) & from 2-D sine fit & from 2-D sine fit& from 2-D sine fit\\
\hline
50 & $(1.413)^c$ & $1.401\,\pm\,(>0.100)^a$ & -- & -- & $1.520\,\pm\,(>0.053)^a$ & -- \\
\hline
180 & $1.485\,\pm\,0.018$ & $1.491\,\pm\,0.009$ & 1.482 & $1.493\,\pm\,0.005$ & $1.438\,\pm\,0.004$ & $1.548\,\pm\,0.003$ \\
 & $(1.562)^c$ & $1.547\,\pm\,0.015$ & -- & & &\\
\hline
196 & $(1.460)^c$ & $1.455\,\pm\,0.010$ & 1.459 & $1.496\,\pm\,0.006$ & $1.468\,\pm\,0.005$ & $1.524\,\pm\,0.004$  \\
 & $(1.502)^c$ & $1.506\,\pm\,0.008$ & 1.504 &&&\\
\hline
302 & $(1.483)^c$ & $1.478\,\pm\,0.044$ & -- & -- & $1.488\,\pm\,0.012$ & -- \\
\hline
371 & -- & -- & -- & -- & $1.442\,\pm\,0.011$ & --\\
\hline
526 & $(1.487)^c$ & -- & -- & -- & $1.452\,\pm\,0.028$ & --\\
\hline
742 & $(1.461)^c$ & $1.461\,\pm\,(>0.016)^b$ & -- & -- & $1.498\,\pm\,0.011$ & --\\
 & $(1.524)^c$ & $1.522\,\pm\,(>0.017)^b$ & -- &&&\\
\hline
1286 & $(1.489)^c$ & $1.489\,\pm\,0.037$ & -- & -- & $1.496\,\pm\,0.011$ & --\\
\hline
\end{tabular}
\end{center}
$^a$unconstrained on 3-$\sigma$ level and only 1-$\sigma$ level given\\
$^b$unconstrained on 3-$\sigma$ level and only 2-$\sigma$ level given\\
$^c$unconstrained on 1-$\sigma$ level and only the nominal value is given
\label{det_overtones}
\end{table*}

The detected overtone frequencies listed in Table~\ref{det_overtones}
need to be compared to the expected overtone frequencies based on the
measured ground harmonics listed in Table~\ref{det_signals}, which
are given in the last three columns, assuming the results from the
2-D sine fitting listed in the second column of Table~\ref{det_signals}.
Expected overtones are those from
$f_1 + f_2$, $2 \times f_1$, and $2 \times f_2$. Within the combined
uncertainties, all observed overtone frequencies can be uniquely
associated to one of the expected overtone frequencies.
The overtones determined from using the Lomb-Scargle algorithm
are also in satisfactory agreement within the larger uncertainties.
These findings are also in agreement with \cite{leibowitz2006}.

\subsection{The X-ray spectra}
\label{spectra}

\begin{figure}
\includegraphics[width=84mm]{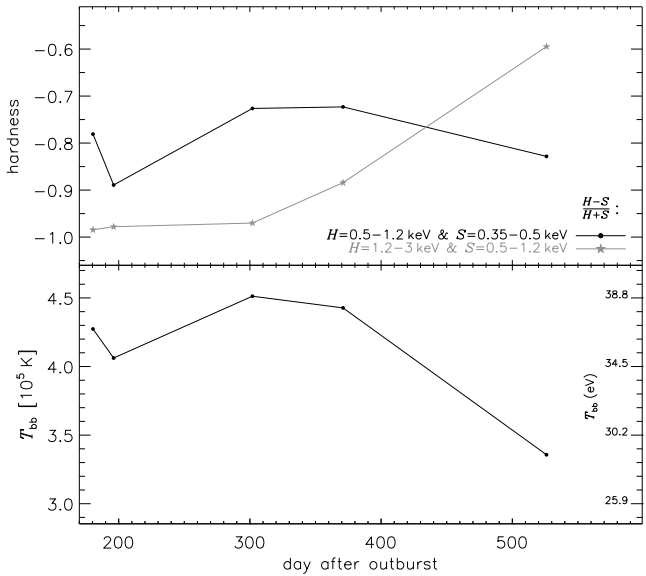}
\caption{\label{bevol}Parameterisation of the spectral evolution during the
SSS phase.
In the top panel, spectral hardness ratios with two different energy cuts,
as detailed in the bottom right legend, are shown. In the bottom panel,
best-fit colour temperatures derived from blackbody fits are shown, that
show the same evolution as the hardness ratio curve plotted in black.}
\end{figure}

 On order to compare our photometric results with the spectral evolution,
we have extracted the high-resolution X-ray grating spectra taken during
the SSS phase, thus on days 180, 196, 302, 371, and 526, using standard
data reduction packages provided by the \chandra\ and \xmm\
missions. For details on spectral analyses, we refer to, e.g.,
\cite{ness2007}. The SSS spectrum taken on day 180 has
been shown in fig.~3 of \cite{ness2003}. We have also extracted
the low-resolution spectra taken on day 50 with \chandra/ACIS and
on days 742 and 1286 with \xmm/MOS1.

 From the spectra, we have determined X-ray band fluxes over the
energy range 0.3-2.5\,keV, which is covered by all instruments.
For the high-resolution grating spectra, the fluxes can be obtained
directly by integrating the photon energies from each spectral
bin. The low-resolution spectra suffer from photon redistribution,
but reliable fluxes can be obtained by fitting a spectral model to
the data that is folded through the instrumental response matrix.
The evolution of X-ray fluxes is shown in Fig.~\ref{aevol}.

For the SSS spectra, we have determined
the spectral colour for each spectrum, following two approaches. First,
we computed two different hardness ratios $HR=(H-S)/(H+S)$ were $H$ and
$S$ are the number of counts within a hard and a soft energy band,
respectively. We computed two different hardness ratios from different
energy cuts. The first hardness ratio has been computed with
$0.35-0.5$\,keV and $0.5-1.2$\,keV for
soft and hard bands, respectively, and a second one using the two bands
$0.5-1.2$\,keV and $1.2-3$\,keV. In the top panel of Fig.~\ref{bevol},
the evolution of the two hardness ratios is shown using a black and
and a grey colour, as indicated in the bottom right legend. Clearly,
the spectral evolution depends on the choice of the energy ranges.

 We have therefore followed a second approach. As can be seen from fig.~3
in \cite{ness2003}, the shape of the SSS spectrum resembles that of an
absorbed blackbody. Also other SSS spectra have a blackbody-like
shape (see, e.g., fig.~4 in \citealt{ness2008} or fig.~4 in \citealt{ness2007}). We have fitted absorbed blackbody models to each spectrum
and obtained blackbody temperatures. While this quantity has the unit
of a temperature, the values obtained from the spectral fits are most
likely different from the effective temperature. It appears from
previous work that blackbody temperatures derived from spectral fits
to SSS spectra are systematically lower than the effective temperature,
because the corresponding bolometric luminosity is unreasonably high
(\citealt{krautter1996}). The derived values can therefore not be
considered more than a spectral characterisation. It is, however,
one possibility that spectral changes are due to changes in
the photospheric temperature.
As we are interested in the temperature {\em evolution},
we assume that the evolution of the blackbody
temperature reflects the evolution of the effective temperature. While
this remains to be proven, we use this assumption as our working
hypothesis.

 In the bottom panel of Fig.~\ref{bevol} we show the evolution of
the blackbody temperature. We have calculated statistical uncertainties
accounting for correlations with other parameters, but these errors
are smaller than the plot symbols. We caution, however, that a
$\chi^2_{\rm red}$ goodness criterion indicates poor fits, and
statistical errors are of little use in this situation. The reason
for poor fits is the presence of deep absorption lines that are
not reproduced by a blackbody. Since the continuum is well reproduced
in each case, we still consider the blackbody fits a good approximation
for the spectral shape. The best-fit blackbody temperatures are listed
in Table~\ref{tevol}. From day 180 to 196, the
blackbody temperature decreases and recovers to a somewhat higher
value on day 302. The temperature then stays approximately constant
until day 371, after which time it decreases by a large amount.
This behaviour resembles that of the evolution of the softer hardness
ratio shown with the black line in the top panel of Fig.~\ref{bevol}.
The temperature decrease between days 180 and 196 may be associated
to the steep decay on day 180, but it could also be an instrumental effect,
as the spectrum on day 180 was taken with \chandra\ while that
on day 196 was taken with \xmm.

The normalisation can be converted to a radius assuming
spherical geometry. With a distance of $d=3.9$\,kpc, we obtain
values ranging around 100,000\,km. The bolometric luminosities that
correspond to the derived temperatures and radii are two orders above
the Eddington luminosity and must be considered unrealistically large.
Consequently, the derived radius of $\sim 100,000$\,km is overestimated
and can be considered as upper limit.

\section{Discussion}
\label{discussion}

\begin{figure}
\includegraphics[width=84mm]{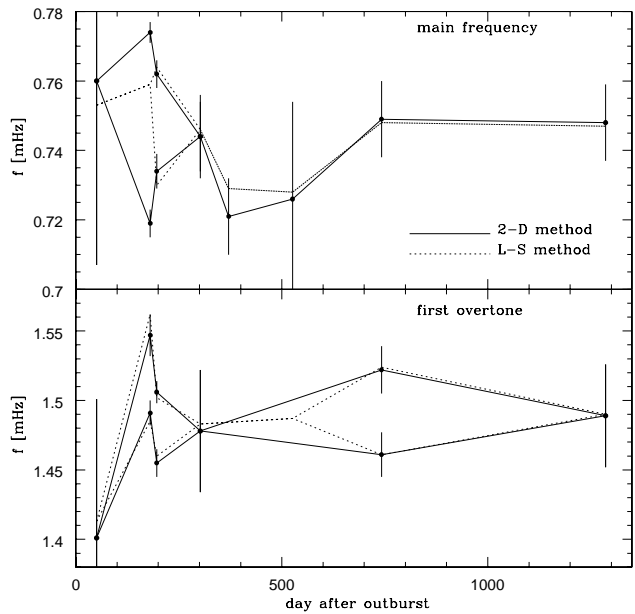}
\caption{Graphical illustration of the evolution of the measured X-ray frequencies.
The top panel shows the main frequency around 0.75\,mHz, and in
the bottom panel we show the first overtone. The bullets, connected by
the solid line are the results from the 2-D method. The results from the
Lomb-Scargle method are included with the dashed lines for comparison.}
\label{pevol}
\end{figure}

With a data set covering more than three years of evolution, we see
large changes in brightness between observations. On day 180, a sharp
decline was observed by \cite{ness2003}, and brightness variations
thus occur on long and short time scales. The origin of the observed
emission is not the same in all observations. On day 50, the X-ray
emission likely originated
from shocks within the ejecta, while on days 180-526, the spectrum was
dominated by SSS emission that comes from the photosphere around the
white dwarf. The post-SSS phase in novae is usually dominated by
optically thin X-ray emission originating from the nebular ejecta that
are radiatively cooling like in V382\,Vel (see \citealt{ness2005}).
V4743\,Sgr has a quiescent X-ray emission source (\citealt{ness2007a})
that could resemble those typically observed in intermediate polars,
where X-ray emission
originates from an accretion shock close to the white dwarf.
In addition, \cite{kang2006} detected a similar frequency of
$0.7047\pm 0.0009$\,mHz in optical observations which is likely
associated with magnetically controlled accretion columns. The authors interpretted this frequency as the beating between the spin frequency of the white dwarf and the orbital modulation. Furthermore our optical/UV detection of a frequency of $0.699 \pm 0.015$\,mHz during day 742 is consistent with the \cite{kang2006} optical finding within the errors.
It thus stands to
reason that the origin of the main X-ray frequency is of a fundamental nature
as, e.g., the spin period of the
white dwarf. However, systematic inspection of the frequency evolution
yields some inconsistencies with the interpretation of pure rotational
modulation.

We confirm the multifrequency substructure found by \cite{leibowitz2006} for days 180 and 196. However, for day 180, we need the
higher sensitivity of our 2-D sine fitting method, as the
Lomb-Scargle method yields only a single peak. Similarly,
Leibowitz et al. had to use a least square fitting approach in order
to detect the double nature of the main peak for day 180.
Since in addition, the formal measurement uncertainties from the
Lomb-Scargle method are higher than for our 2-D sine fitting method,
we concentrate only on the results from the 2-D sine fitting method
for the discussion.

\subsection{Evolution of the main frequency}

 The evolution of the main frequency is illustrated in
Fig.~\ref{pevol}. In the early observations, it is split
in two frequencies, as the two signals on days 180 and 196 are
different at $<99.7$\% confidence. Since they have
moved closer together from day 180 to day 196, it is possible that
this was a monotonic trend that continued until they merged into
a single frequency. Since the exposure times on days 301, 372,
and 526 were much shorter than the beat period that corresponds
to the difference between the two signals on day 180, this possibility
can not be tested. However, as illustrated in Fig.~\ref{beat},
the observations on days 742 and 1286 were long enough to detect
two frequencies if they were as far apart as on day 180. Unless the
relative amplitudes of the two signals have changed significantly,
we can confidently conclude for at least these two observations
that
only a single signal was present and that the double nature of the
0.75-mHz signal has disappeared. These results are also supported by
the overtone studies. The detected values satisfy the characteristics
of a fundamental and a first harmonic.

 After day 196, the frequency appears to have decreased and then
increased again by day 742 (see Fig.~\ref{pevol}). While this is formally
not statistically significant in the 3-$\sigma$ level, other nearby
signals have undergone
such kind of change on a statistically significant level (see
Table~\ref{tab:othersigs}). Furthermore,
the decrease of the relative amplitude
as illustrated in Fig.~\ref{aevol} is highly significant. In addition
to being
split in two nearby frequencies in the early evolution, the main
frequency is thus not a stable signal.

 In spite of the likely different origin of the optical emission
observed by \cite{kang2006}, they found a similar period to our main
frequency, yielding $0.7047\,\pm\,0.0009$\,mHz, which is consistent
with our optical/UV measurements obtained with \xmm\ OM on day
742. While their frequency is remarkably close to our main X-ray
frequency, the statistical uncertainty
ranges indicate a significant difference from all our measurements (see
Table~\ref{det_signals}), including the late observations up to
1286 days after outburst. Nevertheless, the similarity of
the frequency is striking and suggests some fundamentally common
underlying origin. However, if the spin period of the white dwarf
were to be responsible
for the modulations in X-ray and optical, then the
frequencies in both bands should be identical.

The fact that the optical and the X-ray frequencies differ, plus the
obvious changes in frequency and amplitude, demonstrate that there is
more than simply rotational modulation from the spin of the white dwarf.

\subsection{Pulsations versus Spin Modulation}
\label{disc:pulsations}

\begin{figure}
\includegraphics[width=84mm]{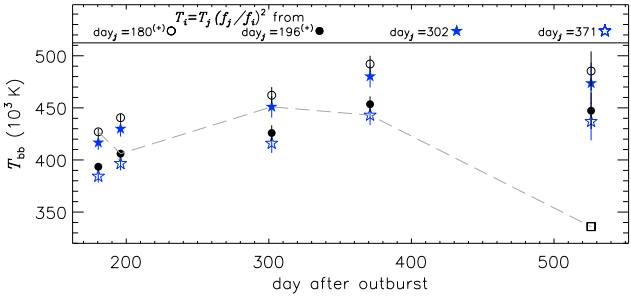} 
\caption{Graphical illustration of the connection between changes in
X-ray frequency and colour temperature. Each data point represents a
predicted temperature $T_i$ for each observation taken at time day$_i$
after outburst (x-axis), according to the formula given in the top,
where $T_j$ is the blackbody temperature (see Sect.~\ref{spectra})
for day$_j$ as indicated by the plot symbol (see legend), and $f_j/f_i$
is the ratio of frequencies from our measurements. The dashed grey line
represents the observed colour temperatures (see Fig.~\ref{bevol})
for comparison with the
predictions (the symbols that are connected by this line are the
cases $i=j$, thus the measured temperatures).}
\label{pulse}
\end{figure}

 While \cite{kang2006} rule out pulsations, the new evidence
from our analysis
requires us to reopen the case. Kang et al. argued that the
interpretation of this period as white dwarf pulsations would
yield a significant slow-down with time, owing to cooling of the
temperature of the white dwarf (\citealt{somers1999}).
From their estimates of changes in effective temperature, derived from
observed luminosity changes, they calculated that the period reported
by \cite{ness2003} for day 180 should be 38 minutes around day
1000, and they rule out pulsations because no such period is
in their data. While it is true that we are also not observing a 38
minute period in the X-ray data around day 1000, we do see significant
frequency changes in Fig.~\ref{pevol}.

\cite{kang2006} suggested using equation 10 in \cite{kjeldsen1995} for the conversion of temperature changes to expected pulsation
frequency changes. In Sect. 4.2 of  \cite{kjeldsen1995}, a
scaling relation between stellar parameters and pulsation frequencies
is derived. It is argued that the acoustic cut off frequency,
$\nu_{\rm max}$, that defines a typical time scale of the
atmosphere, scales with the maximum envelope frequency of stellar
pulsations. Linear adiabatic theory is applied to derive $\nu_{\rm max}$
and thus the relation given in their equation 10. The ratio of two
pulsation frequencies in two different atmospheres scales with the
square root of the inverse ratio of the respective effective
temperatures (assuming the same mass and radius). With this method,
our observed changes in frequency can be converted to expected changes
in photospheric temperature, if pulsations were assumed. In
Table~\ref{tevol} we list squared ratios
of frequencies from our 2-D sine fitting method taken from
Table~\ref{det_signals}. We indicate the two respective observation
dates with numbers (1) and (2) for which the ratios were taken.
For days 180 and 196 we indicate the high and low frequencies by
superscript '+' and '-' behind the day of observation. Since the
3-$\sigma$ uncertainties given in Table~\ref{tevol} do not include
the possibility that two unresolved frequencies are present, we
used the larger uncertainty range resulting from half of the inverse
exposure times for days 302, 371, and 526. The inverse exposure time
sets a strict limit on the detectability of two separate frequencies
(see Sect.~\ref{sect:beat} and Fig.~\ref{beat}),
and if the signal was split in two components on days 302, 371, and
526, then a value somewhere in between would result (depending on
the individual amplitudes), while the
statistical uncertainty would not include the upper or lower frequency.

 As an estimate of the evolution of the effective temperature, we use
colour temperatures derived in Sect.~\ref{spectra}. As pointed out
earlier, the blackbody temperature is not equivalent to the photospheric
temperature, and also the evolution might not follow the same trend.
However, we consider it a strong possibility that changes in the
blackbody temperature indicate equivalent changes in temperature
and assume this as a working hypothesis. The following conclusions
depend on whether or not this assumption is valid.

 In the last three columns of Table~\ref{tevol} we quote the colour
temperatures derived in Sect.~\ref{spectra} for the respective
days and the inverse ratios of these values for direct comparison
with the squared frequency ratios in the third column. The extent
of agreement between these ratios is an indicator for the validity
of the underlying assumption of pulsations. Those ratios which agree
within the given uncertainty ranges of the frequency measurement are
indicated by underlined numbers. For illustration we show in
Fig.~\ref{pulse} the evolution of the observed colour temperature
in comparison with temperatures computed from
\begin{equation}
T_i=T_{j}*(f_j/f_i)^2\,,
\end{equation}
where $T_i$ and $f_i$ are predicted temperature and observed frequency
for day$_i$, respectively, and $T_j$ and $f_j$ are the observed
temperature and frequency for day$_j$.

 The best agreement between observed and predicted effective
temperature is seen for the observations on days 180 and 302, and
between days 196 and 371. All predictions derived from day 526 are
far off the mark and are not included in the plot. This can be seen
from the predictions from other observations for day 526, which are
too high, indicating that the
frequency changes do not predict the decline in temperature.
 The agreement between predicted and observed temperature
between days 196 and 302 and between days 180 and 196 is
poorer. In light of the underlying assumptions made by
\cite{kjeldsen1995}, a main sequence (MS) stars with
an effective temperature of order 5500\,K, plus the
considerable uncertainty in the effective temperature,
the agreements are surprisingly good, strongly suggesting
that pulsations play an important role. The breakdown of
the relationship towards the end of the SSS phase can be
understood as the turn off of the central energy source,
leading to a highly non-equilibrium situation, and pulsations
may not propagate the same way as in equilibrium. We thus
argue that the possibility of pulsations can at least not
easily be ruled out.

 It must be noted that for a frequency of $\sim 0.75$\,mHz
and a mass of order 0.5-1.4\,M$_\odot$, equation 10
of \cite{kjeldsen1995} yields effective radii of order
2-7\,R$_\odot$, which appears rather large for a hot WD
and is about an order of magnitude larger than the radii
derived in Sect.~\ref{spectra} from blackbody fits, which
have been discussed to be unrealistically large.
If this apparent discrepancy could be resolved by a larger
scaling factor, then the relative changes of effective
temperature and oscillation changes would still hold for
a hot white dwarf. Without computing new models for hot white
dwarfs, the validity of the relation must be taken with care.
However, if we assume this relation to be valid,
then we have good reason to conclude that pulsations are
occurring between days 180 and 371.\\

 A fact that complicates the interpretation is the
similarity between the frequencies measured during the SSS
phase and during the last phase, long after the nova has
turned off. This similarity suggests that a fundamentally
similar origin is accountable, which brings us back to
the spin period of the white dwarf. While pulsations
seem to be a valid explanation for the SSS phase, the spin
period of the white dwarf should also play a major role.
We have no explanation but hesitate to believe in a
coincidence. Perhaps magnetic fields that are present
in an intermediate polar might stir up the ejecta during the
SSS phase and stimulate pulsations. However, this is pure
speculation.

\cite{leibowitz2006} interpreted the 0.75\,mHz structure as due to
the rotation of the white dwarf and other close signals as the nonradial
pulsations of the central star. This explanation would be consistent with
a similar interpretation by \cite{drake2003} and \cite{rohrbach2009}
for the nova V1494\,Aql. A dominant short-term oscillation with a period
of 2498.9\,s with other close signals was detected. The authors interpreted
the variability as the pulsations of the central white dwarf. These central
acretors after the nova explosion resemble planetary nebula nuclei where
pulsations were observed in the range of $\sim 1000$\,s to $\sim 5000$\,s
(\citealt{ciardullo1996}). For example, the white dwarf
in the planetary nebula NGC\,246 had a principal frequency of 0.67\,mHz
in September 1989 and 0.54\,mHz in June 1990 (\citealt{ciardullo1996}).
This gives a change of 0.13\,mHz within 9 months, i.e., $\sim 0.01$\,mHz
per month. This is remarkably similar to our finding of changes in each of
the two frequencies detected on days 180 and 196 (Table~\ref{det_signals}),
which change by the amount of $\sim 0.01$\,mHz within two weeks.
Such frequency changes are thus not unprecedented in pulsating white dwarfs.

\cite{sastri1973} studied instabilities in hydrogen burning nova
shells and found pulsations as a possible phenomenon. However, while the
multiperiodic and unstable behaviour is consistent with our results, the
predicted typical periods are much smaller. For example, the unstable
35\,s X-ray oscillations in RS\,Oph (\citealt{beardmore2008}) shortly after
the start of the super soft phase would fall into this range.

Another mechanism with periods in accordance to our finding
could be pulsations in isolated hot GW\,Vir white dwarfs. The origin of these
pulsations is the compression-induced opacity and ionisation increase in
the partial ionisation zones of carbon and oxygen (\citealt{starrfield1984}).
Such nonradial pulsations are typical for hydrogen-deficient white dwarfs
with carbon-oxygen envelopes. While in accreting white dwarfs like
V4743\,Sgr the white dwarf is not hydrogen poor on its surface,
\cite{dreizler1996} found some GW\,Vir
stars that have pulsations and hydrogen on their surface.
The X-ray spectra of V4743\,Sgr contain relatively deep carbon and oxygen
absorption lines (\citealt{ness2003}), which is an indication for
V4743\,Sgr being a CO-type nova as opposed to an ONe-type nova.
\cite{drake2003} proposed the same interpretation in the case of the nova V1494\,Aql, but without any spectral indication of the white dwarf to be the required CO type.

\begin{table}
\caption{Comparison of squared X-ray frequency ratios (frequencies from
2-D method from Table~\ref{det_signals}) with inverse
ratios of colour temperature ($T_{\rm col}$ in
$10^3$\,K, derived in Sect.~\ref{spectra}. For days 180 and 196,
the low and high frequencies are marked with - and + superscripts,
respectively. The errors include systematic uncertainties from
the exposure times (see text for details).
The underlined numbers indicate cases for which the relation
$(f_1/f_2)^2\simeq T_{\rm col}^{(2)}/T_{\rm col}^{(1)}$ holds.
Note that mismatches involve only days 196 and 526, times of
non-equilibrium situations (see Sect.~\ref{disc:pulsations}).}
\begin{center}
\begin{tabular}{lccccc}
\hline
\hline
Day(1) & Day(2) & $(f_1/f_2)^2$ &  $T_{\rm col}^{(1)}$ & $T_{\rm col}^{(2)}$ & $T_{\rm col}^{(2)}/T_{\rm col}^{(1)}$\\
\hline
180$^-$ & 196$^-$ & \underline{0.96}$\,\pm\,0.01$ & 427 & 406 & \underline{0.95}\\
180$^+$ & 196$^+$ & $1.03\,\pm\,0.01$ & 427 & 406 & 0.95\\
196$^-$ & 302 & $0.97\,\pm\,0.11$ & 406 & 451 & 1.11\\
196$^+$ & 302 & \underline{1.05}$\,\pm\,0.11$ & 406 & 451 & \underline{1.11}\\
302 & 371 & \underline{1.06}$\,\pm\,0.16$ & 451 & 443 & \underline{0.98}\\
371 & 526 & $0.99\,\pm\,0.17$ & 443 & 336 & 0.76\\
180$^+$ & 302 & \underline{1.08}$\,\pm\,0.11$ & 427 & 451 & \underline{1.06}\\
180$^+$ & 371 & \underline{1.15}$\,\pm\,0.12$ & 427 & 443 & \underline{1.04}\\
180$^+$ & 526 & $1.14\,\pm\,0.13$ & 427 & 336 & 0.79\\
302 & 526 & $1.05\,\pm\,0.17$ & 451 & 336 & 0.74\\
\hline
\end{tabular}
\end{center}
\label{tevol}
\end{table}

\subsection{Connection between spin and orbital period}

\cite{leibowitz2006} and \cite{kang2006} argued that the two
nearby frequencies can arise from the white dwarf spin period on
the one hand and the beat period between the spin and the
orbital period of the binary system on the other hand.
The orbital period was reliably determined by \cite{kang2006}
as $6.718\,\pm\,0.005$ hours from both observing campaigns.
This can be compared to the beat periods between various pairs
of observed frequencies. In Table~\ref{periods} we list the detected
periods in minutes, the difference between their frequencies, and the
derived beat periods. We calculated the beat periods for the two
signals in the X-ray light curves taken on days 180 and 196 (top
two rows), for all possible combinations between these two signals
and the optical period of 23.651\,min
(derived from the $f_1=0.7047$\,mHz frequency found
by \cite{kang2006}; following four rows), and the
combinations of periods observed in the late X-ray light curves
and the optical period (bottom two rows).

The beat periods between the two periods detected on days 180
and 196 are clearly inconsistent with the orbital period. Also,
the beat periods between these frequencies and the optical frequency
are not consistent with the orbital period. Only the beat periods
between the frequencies derived from the late X-ray observations
and the optical frequency are in agreement with the orbital
period. The latter agreement suggests that either the optical
or the late X-ray period is the spin period, while the respective
other one is the beat between orbital and spin period. The most probable situation is that the X-ray signal is the spin of the white dwarf.
The fact that none of the earlier frequencies are connected to the orbit-spin
relation, lends additional support to our interpretation that the
oscillations during the SSS phase are more than only the spin
period of the white dwarf.

\begin{table}
\caption{Beat periods $P_{\rm beat}$ (last column in unit hours), derived
from various pairs of observed periods '$P_1$' and '$P_2$' (in unit
minutes) and their corresponding frequencies, $f_1$ and $f_2$.
In the top two rows, the beat periods for the
two close frequencies that we found in the X-ray light curves on days
180 and 196 (see Table~\ref{det_signals}) are given. Next, the
beat period between each X-ray period and optical period of 23.651
minutes, observed between days 1003 and 1011 by Kang et al. (2006),
are given. The last two rows show the beat period between the late X-ray
periods and the optical period, which compare well to the orbital
period of 6.72 hours.}
\begin{center}
\begin{tabular}{lccrr}
\hline
\hline
Day & $P_1$ [min] & $P_2$ [min] & $f_2$ - $f_1$ [mHz] & $P_{\rm beat}$ [h]\\
\hline
180 & $23.180$ & $21.533$ & $0.055\,\pm\,0.005$ & $5.1\,\pm\,0.5$\\
196 & $22.707$ & $21.872$ & $0.028\,\pm\,0.006$ & $9.9\,\pm\,2.2$\\
\hline
180 & $23.180$ & $23.652^a$ & $0.069\,\pm\,0.003$ & $4.0\,\pm\,0.2$\\
180 & $21.533$ & $23.652^a$ & $0.014\,\pm\,0.004$ & $19.8\,\pm\,6.2$\\
196 & $22.707$ & $23.652^a$ & $0.057\,\pm\,0.004$ & $4.9\,\pm\,0.3$\\
196 & $21.872$ & $23.652^a$ & $0.029\,\pm\,0.005$ & $9.6\,\pm\,1.7$\\
\hline
742 & $22.252$ & $23.652^a$ & $0.044\,\pm\,0.011$ & $6.3\,\pm\,1.7$\\
1286 & $22.282$ & $23.652^a$ & $0.043\,\pm\,0.011$ & $6.4\,\pm\,1.8$\\
\hline
\end{tabular}
\end{center}
$^a$Detected optical period between days 1003 and 1011 ($f_2=0.7047 \pm 0.0009$\,mHz)
\label{periods}
\end{table}

\subsection{The coincidence between pulsation and spin modulation}

 In Sect.~\ref{disc:pulsations} we have discussed that during the
SSS phase, the main period could be characterised by pulsations, while
the frequency detected in the later observations is more likely the
spin period of the white dwarf. In that case we wonder why the
frequencies are so similar, despite of their different nature
(see, e.g., Fig.~\ref{LSpower_all_run}).

 We consider it unlikely that the spin period of the white dwarf has
changed, yet small but significant changes in the observed oscillation
frequency are undeniable. We can only speculate about how to explain
the presence of a persistent signal that is variable at the same time.
Perhaps, the spin period of the white dwarf could be modulated by additional
processes that could depend on the physical conditions of the emitting
plasma. One possibility would be that the magnetic field axis is not
aligned with the rotation axis, and the spinning of the non-aligned
magnetic dipole stirs up the ejecta surrounding the white dwarf.
In this way pulsations could be stimulated by the spin period of the
white dwarf. The flux is then modulated by these pulsations and the time evolution would depend on the properties of
the ejecta. We also note that the expansion of the shell during the SSS phase that
rotates at a slower velocity in the outer layers, owing to conservation
of angular momentum, could slow down the oscillation frequency. We
caution, however, that these ideas are unsubstantiated without
theoretical support and are as such highly speculative.

\subsection{The decline on day 180}

 In Figs.~\ref{aevol} and \ref{pevol} it can be seen that the
largest changes in frequency and amplitude as well as the shrinking
difference between the two components of the main peak appeared shortly
after the sudden brightness decline on day 180 that was accompanied
by spectral changes from soft optically thick to hard optically thin
(\citealt{ness2003}). Since this decline has only been seen in the
first SSS observation, thus during the
early SSS phase, it may be part of an early variability phase similar
to that first seen by Swift in RS Oph (e.g., \citealt{beardmore2008}). Such
an early variability phase is now routinely being observed, e.g., in
V458\,Vul (see figure 6
in \citealt{ness2009}). Since these novae did not show any persistent
oscillations as in V4743\,Sgr, our findings are a unique contribution
to the discussion about the origin of the early variations in the
SSS phase. Additional evidence comes from high-resolution spectroscopy.
Ness (2010, in preparation) reports the rapid disappearance of some of the nebular
emission lines that were seen during the dark phase (\citealt{ness2003}).
This indicates that the surrounding nebular emission was recombining
after the bright central X-ray source had disappeared. One possibility
for the complete disappearance of the X-ray source could be an expansion
of the photospheric radius, followed by photospheric adjustment, thus
shifting the peak of the SED back into the ultraviolet. This would be
consistent with the spectral changes parameterised in Fig.~\ref{bevol}.
From day 180 to day 196, a significant drop in the blackbody temperature
and hardness is seen which could indicate a lower photospheric
temperature on day 196. In that case,
the atmospheric structure could change significantly, which could lead
to changes in pulsation frequency and amplitude, as the conditions for
the propagation of pulsations through the atmosphere would change.
Our observations of changes in the oscillations are important evidence
for a physical explanation of the variations rather than geometrical
causes such as an eclipse.

\section{Summary}
\label{summary}

In this paper, all available X-ray light curves of the nova V4743\,Sgr
are presented and analysed. Our results have to be viewed in light of
three different phases of evolution, each with
a different origin of X-ray emission:
\begin{itemize}
\item Early hard X-ray emission phase with X-rays likely originating from
shocks within the ejecta. This phase is characterised by weak emission
line spectra that can be modeled with optically thin thermal models.
Our observation taken on day 50 belongs to this phase.
\item Super Soft Source (SSS) phase with the X-ray emission
originating from a pseudo photosphere around the white dwarf.
This phase is characterised by extremely bright continuum emission.
The broad-band spectrum can be fitted by a blackbody, and high-resolution
X-ray spectra show absorption lines that are blue shifted (see
\citealt{ness2009}). Our observations taken on days 180, 196, 302, 371, and
526 belong to this phase.
\item Quiescent phase. In Classical novae, the radiatively cooling
ejecta can be observed in X-rays. The spectra from days 742 and 1286 
resemble more those of quiescent intermediate polars (Ness et al.
in prep).
\end{itemize}

Despite of all the differences in
the emission origin, the same main frequency was observed at all
times. In addition, a similar frequency was observed in the optical
(\citealt{kang2006} and our \xmm\ optical/UV OM data), although the
optical light has a different origin. If the optical period is interpreted
as the beat period between the WD spin and the orbit, then the optical
light must originate from something fixed in the orbiting system,
for example from the secondary star or the disc-stream impact region. 
Not many periodic processes in a cataclysmic variable can
leave their footprint in all forms of light, leading us to the
conclusion that the spin period of the white dwarf must play a
major role.

 Meanwhile, the different phases can lead to different kinds
of modulation of the main frequency, and we have investigated
small anomalies that require detailed analysis. Our main results
are:

\begin{itemize}
\item On days 180 and 196, the main signal consists of two nearby
frequencies.
The difference between these two components shrinks from
day 180 to day 196. While only a single frequency was found for
days 302 to 526, these observations were too short to detect such
close signals. Either, the trend of a shrinking difference has
continued until it merged into a single signal, or roughly the
same difference has remained through the SSS phase.
\item The observations of days 742 and 1286 contained only a single
signal, and the double nature has not continued into the quiescent
phase.
\item The average frequency decreased slightly after day 196 and
then increased again after day 526. Colour temperatures derived
in Sect.~\ref{spectra} from fits to the SSS spectra on days 180, 302,
and 371, seem to change in relation to the changes in frequency.
This suggests pulsations as an origin for the modulations.
\item
The relative amplitude of the oscillations decreased after day 196
and then increased again to the original value after day 526.
\item
The largest change in frequency and oscillation amplitude occurred
after the sudden brightness decline on day 180 and could thus be
connected to this event.
\item
We found overtones at $\sim 1.5$\,mHz in the periodograms of
all observations, which
are first harmonics and combinations of two close fundamentals
present near the main feature at $\sim 0.75$\,mHz.
\item
The beat period between the X-ray period and the optical period
during the quiescent phase coincides with the orbital period.
\end{itemize}
 We believe that during the SSS phase, pulsations are the major source
for the oscillations, while the oscillations seen during the quiescent
phase are the spin period of the white dwarf. The similarity of the
spin period and pulsation period is puzzling, and no mechanism that
could co-align the periods is known to us. Based on the spin period,
we support the interpretation by \cite{kang2006} that V4743\,Sgr is
an intermediate polar.

\section*{Acknowledgement}

AD acknowledges the Slovak Academy of Sciences Grant No. 2/0078/10. We thank V.Antonuccio-Delogu from INAF Catania for providing computing facilities.
We acknowledge financial support from the Faculty of the European Space Astronomy Centre.

\vspace{-.6cm}

\bibliographystyle{mn2e}
\bibliography{mybib}

\label{lastpage}

\end{document}